\documentclass[%
aps,prb,superscriptaddress,twocolumn,floatfix,10pt]{revtex4-2}
\usepackage[utf8]{inputenc}
\usepackage{amsmath,amssymb}
\usepackage{empheq}
\usepackage{float}
\usepackage{lipsum}
\usepackage{mathtools,cuted}
\usepackage{esvect}
\usepackage{multirow}
\usepackage{xcolor}
\usepackage{soul}
\usepackage[export]{adjustbox}
\usepackage{graphicx}
\usepackage{dcolumn}
\usepackage{bm}
\usepackage{hyperref}
\usepackage{physics}
\usepackage[mathlines]{lineno}
\usepackage[capitalise]{cleveref}
\usepackage{todonotes}
\usepackage{bbm}
\usepackage{orcidlink}

\newcommand{\ham}[0]{H}
\newcommand{\sbar}[0]{\overline{S}}

\begin{document}
	
	\title{Emergent non-Hermitian models}
	
 \author{Lumen Eek\,\orcidlink{0009-0009-1233-4378}}
 \affiliation{Institute of Theoretical Physics, Utrecht University, Utrecht, 3584 CC, Netherlands}
 \author{Anouar Moustaj\,\orcidlink{0000-0002-9844-2987}}
 \affiliation{Institute of Theoretical Physics, Utrecht University, Utrecht, 3584 CC, Netherlands}
 \author{Malte Röntgen\,\orcidlink{0000-0001-7784-8104}}
 \affiliation{
 	Laboratoire d’Acoustique de l’Université du Mans, Unite Mixte de Recherche 6613, Centre National de la Recherche Scientifique, Avenue O. Messiaen, F-72085 Le Mans Cedex 9, France
 }
 \author{Vincent Pagneux\,\orcidlink{0000-0003-2019-823X}}%
 \affiliation{
 	Laboratoire d’Acoustique de l’Université du Mans, Unite Mixte de Recherche 6613, Centre National de la Recherche Scientifique, Avenue O. Messiaen, F-72085 Le Mans Cedex 9, France
 }
  \author{Vassos Achilleos\,\orcidlink{0000-0001-7296-2100}}
 \affiliation{
 	Laboratoire d’Acoustique de l’Université du Mans, Unite Mixte de Recherche 6613, Centre National de la Recherche Scientifique, Avenue O. Messiaen, F-72085 Le Mans Cedex 9, France
 }
   \author{Cristiane Morais Smith\,\orcidlink{0000-0002-4190-3893}}
   \affiliation{Institute of Theoretical Physics, Utrecht University, Utrecht, 3584 CC, Netherlands}

	\date{\today}
	
	\begin{abstract}
    The Hatano-Nelson and the non-Hermitian Su-Schrieffer-Heeger model are paradigmatic examples of non-Hermitian systems that host non-trivial boundary phenomena.
    In this work, we use recently developed graph-theoretical tools to design systems whose isospectral reduction---akin to an effective Hamiltonian---has the form of either of these two models.
    In the reduced version, the couplings and on-site potentials become energy-dependent. We show that this leads to interesting phenomena such as an energy-dependent non-Hermitian skin effect, where eigenstates can simultaneously localize on either ends of the systems, with different localization lengths. Moreover, we predict the existence of various topological edge states, pinned at non-zero energies, with different exponential envelopes, depending on their energy. Overall, our work sheds new light on the nature of topological phases and the non-Hermitian skin effect in one-dimensional systems.
	\end{abstract}
	
	\maketitle{}
	
	
	\section{\label{sec:Intro}Introduction
	}
In recent years, the study of non-Hermitian physics has gained significant attention due to its profound impact on the fields of condensed matter, meta-materials, acoustics, and photonics \cite{McDonald2020Exponentially-enhancedDynamics,San-Jose2016MajoranaSuperconductors,Ghatak2020ObservationMetamaterial,Sone2020ExceptionalMatter}. Indeed, non-Hermitian platforms offer enhanced sensing capabilities \cite{McDonald2020Exponentially-enhancedDynamics}, can exhibit Majorana bound states near exceptional points \cite{San-Jose2016MajoranaSuperconductors}, and provide opportunities for utilizing topological edge modes in the field of active matter \cite{Ghatak2020ObservationMetamaterial,Sone2020ExceptionalMatter}. In addition, the non-conservative and non-unitary dynamics of non-Hermitian systems have led to the discovery of phenomena that challenge the conventional notions of symmetry and stability \cite{Kunst2018PRL121026808BiorthogonalBulkBoundaryCorrespondenceNonHermitian,Bergholtz2021ExceptionalSystems,Kawabata2019PRX9041015SymmetryTopologyNonHermitianPhysics,Gong2018PRX8031079TopologicalPhasesNonHermitianSystems}. Among these, the non-Hermitian skin effect (NHSE), manifesting itself as an accumulation of modes at the boundaries of the system, has been intensely studied in the past few years \cite{Lee2016AnomalousLattice,Yao2018PRL121086803EdgeStatesTopologicalInvariants,Okuma2020TopologicalEffects,Zhang2021NC126297AcousticNonHermitianSkinEffect,Liu2021Non-HermitianCircuit,Gu2022NC137668TransientNonHermitianSkinEffect}. The NHSE has been realized in multiple platforms, such as acoustic crystals \cite{Zhang2021NC126297AcousticNonHermitianSkinEffect}, electric circuits \cite{Liu2021Non-HermitianCircuit}, and optical lattices using ultra-cold atoms \cite{Liang2022DynamicAtoms}.
Moreover, the interplay between non-Hermitian physics and topology has given rise to novel topological phases \cite{Bandres2018P11NonHermitianTopologicalSystems,Kawabata2019PRX9041015SymmetryTopologyNonHermitianPhysics,Kunst2019PRB99245116NonHermitianSystemsTopologyTransfermatrix,Li2020PRL124250402TopologicalSwitchNonHermitianSkin}. Indeed, when considering Hamiltonians that are no longer Hermitian \cite{Kawabata2019PRX9041015SymmetryTopologyNonHermitianPhysics,Altland1997NonstandardStructures}, the Altland-Zirnbauer topological classification for non-interacting fermions is enlarged from 10 to 38 classes. Additionally, the bulk-boundary correspondence generally no longer holds, and requires substantial modifications to account for boundary phenomena \cite{Kunst2018PRL121026808BiorthogonalBulkBoundaryCorrespondenceNonHermitian,Yang2020Non-HermitianTheory}.

The investigation of toy models has been instrumental in shaping a theoretical comprehension of non-Hermitian systems.
The Hatano-Nelson and the non-Hermitian Su-Schrieffer-Heeger (NH SSH) models \cite{Hatano1997VortexMechanics,Su1979PRL421698SolitonsPolyacetylene,Lieu2018TopologicalModel}, for instance, have become paradigmatic examples of systems hosting the NHSE and a non Bloch bulk-boundary correspondence, respectively.
Upon departing from these idealized models, the same phenomena may take place, but may be more difficult to describe. One way to bridge this difficulty is to reduce the complicated problem into one described by simpler models, with additional features revealed through the reduction process.

An interesting technique---originally introduced for the analysis of graphs and network models---that may be used for this purpose is the so-called isospectral reduction (ISR) \cite{Bunimovich2014IsospectralTransformationsNewApproach}.
The idea behind the ISR is to reduce the matrix dimensionality whilst preserving the spectrum of the original Hamiltonian $H$. This is achieved by recasting the original linear eigenvalue problem into a non-linear one.
The reduced dimensionality simplifies certain tasks and may, in particular, reveal hidden structures of the system \cite{Bunimovich2019AMNS4231FindingHiddenStructuresHierarchies}.
Pivoting around this favourable properties, the ISR has been applied to different problems, for instance, to yield better eigenvalue approximations \cite{Bunimovich2012LAIA4371429IsospectralGraphReductionsImproved} or to study pseudo-spectra of graphs and matrices \cite{VasquezFernandoGuevara2014NLAA22145PseudospectraIsospectrallyReducedMatrices}. 
In physics, the ISR is often encountered in the form of an effective Hamiltonian. One example of this is the Brillouin-Wigner perturbation theory, where the partitioning is done in terms of degenerate subspaces of an unperturbed Hamiltonian \cite{Brillouin1932LesSelf-consistents}. Another example would be integrating out degrees of freedom, where the partitioning is done in Fock space, or integrating out high momentum modes \cite{Hubbard1959CalculationFunctions}. In that context, the reduction provides a suitable starting point for perturbation theory. 

In the last few years, the ISR has also been applied to uncover hidden---so-called latent---symmetries \cite{Smith2019PA514855HiddenSymmetriesRealTheoretical} \footnote{We note that it has recently been shown \cite{Kempton2020LAIA594226CharacterizingCospectralVerticesIsospectral} that the concept of latent symmetry is, for the special case of latent reflection symmetries, equivalent to the graph-theoretical concept of cospectrality \cite{Schwenk1973PTAAC257AlmostAllTreesAre}. A good overview over the properties of cospectrality is given in Chapter 3 of \cite{Godsil2017A1StronglyCospectralVertices}.}.
Latent symmetries become apparent after reduction and have been studied in a number of applications, including quantum information transfer \cite{Rontgen2020PRA101042304DesigningPrettyGoodState}, the design of lattices with flat bands \cite{Morfonios2021PRB104035105FlatBandsLatentSymmetry} or the explanation of accidental degeneracies \cite{Rontgen2021PRL126180601LatentSymmetryInducedDegeneracies}. Very recently, latent symmetries have also been explored in waveguide networks \cite{Rontgen2023PRL130077201HiddenSymmetriesAcousticWave,Rontgen2023EquireflectionalityCustomizedUnbalancedCoherent}, including a possible application in secure transfer of information \cite{Sol2023CovertScatteringControlMetamaterials}.

In this work, we propose to apply the ISR to a range of one-dimensional (1D) non-Hermitian tight-binding models, such that they reduce to the paradigmatic Hatano-Nelson and the NH SSH models. This method allows us to predict the existence of various topological phases and non-standard NHSE as well as to uncover various properties, which were hitherto still unexplored. As an example of the unusual characteristics of this class of models, our approach reveals that they exhibit an energy (or frequency)-dependent NHSE, where eigenstates can localize on either end of the systems. The degree of localization of the NHSE is also influenced by this energy dependence. A similar behavior was recently observed in a system of coupled ring-resonators \cite{Lin2021OLO463512SteeringNonHermitianSkinModes,Xin2023ManipulatingResonators}, for which we extend the theoretical understanding. In addition, we find various topological states pinned at different non-zero energies, protected by a latent spectral symmetry that is only revealed upon applying the ISR. As a consequence of this energy dependence, their exponential envelopes vary, a feature that is also straightforwardly explained, and predicted, upon using the ISR. Throughout this work, we restrict our attention to systems from which the Hatano-Nelson \cite{Hatano1997VortexMechanics} and the NH SSH \cite{Su1979PRL421698SolitonsPolyacetylene,Lieu2018TopologicalModel} models emerge. It should be noted that one could also engineer other types of systems, from which different models would emerge through the ISR.
Indeed, the special case of an asymmetric, Hermitian system from which the conventional SSH model emerges, has been very recently demonstrated \cite{Rontgen2023LatentSuSchriefferHeegerModels}.
\\

This article is structured as follows: in \cref{Sec: ISR}, we lay down the main tool used for the analysis of our models. That is the ISR, which amounts to the construction of an effective Hamiltonian model that is already well understood. This is used on a minimal example, where we do not a-priori expect non-reciprocity to be present. The ISR allows for an intuitive understanding of the reason why the NHSE would arise in such a setup. In \cref{Sec: Emergent HN model}, we extend our analysis to a slightly more complex case, and apply the ISR to a quasi one-dimensional system, resulting in an ``emergent'' Hatano-Nelson model \cite{Hatano1997VortexMechanics}. In this setup, we are able to predict the existence of an energy-dependent NHSE. In \cref{Sec: SSH Reduc}, we add a connection between the unit cells that leads to an ``emergent'' NH SSH model \cite{Lieu2018TopologicalModel}, from which a full understanding of the topological phases can be drawn. In \cref{Sec: Construction Principles}, we generalize the construction principle for which the analysis done for the previous models can be applied. Finally, in \cref{Sec: Conclusion}, we conclude by summarizing our results. 
	
	\section{Isospectral Reduction} \label{Sec: ISR}
    Given a Hamiltonian $H$, it is possible to partition a choice of basis into a set $S$ and its complement $\overline{S}$, so that $H$ can be written in block-form as
    \begin{equation}
        H \equiv \begin{pmatrix}
            H_{S,S} & H_{S,\overline{S}} \\
            H_{\overline{S}, S} & H_{\overline{S},\overline{S}}
        \end{pmatrix} \,.
    \end{equation}
    By partitioning the eigenvalue problem $H\ket{\psi} = E\ket{\psi}$ [where $\ket{\psi}\equiv (\ket{\psi_S}, \ket{\psi_{\overline{S}}})^T$] into the different subsets $S$ and $\overline{S}$ and subsequently eliminating $\ket{\psi_{\sbar}}$, we obtain the non-linear eigenvalue problem
    \begin{equation}
    	\mathcal{R}_S(E,H)  \,, \ket{\psi_S}= E \ket{\psi_S} \,.
    \end{equation}
    Here,
    \begin{equation}
        \mathcal{R}_S(E,H) = H_{S,S} - H_{S,\overline{S}} \left( H_{\overline{S},\overline{S}} -E\mathbbm{1}\right) ^{-1}H_{\overline{S},S} \,,
    \end{equation}
  	is the effective Hamiltonian for the subsystem $S$.
    In the language of graph theory, $\mathcal{R}_S(E,H)$ is known as the ISR of $H$ to $S$ \cite{Bunimovich2014IsospectralTransformationsNewApproach}. An overview of the application of the ISR to physical systems is given in Ref.~\cite{Rontgen2022LatentSymmetriesIntroduction}.
    
     In this work, we show how the ISR allows us to understand the behavior of systems without making approximations. This is done by recognizing that the ISR of a system may yield another (energy-dependent) known model, which is well understood. In this case, the properties of the known reduced system can be used to make predictions for the full system. 
     We shall now illustrate this by means of a simple but important example.\\
\begin{figure}[!hbt]
    \centering
    \includegraphics[width=\columnwidth]{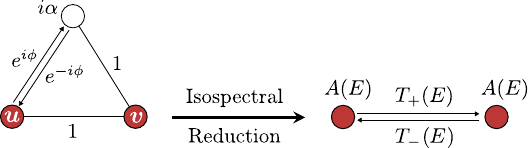}
    \caption{Isospectral reduction of the lossy, complex hopping model on the left to the red sites $S=\{u,v\}$ yields the non-reciprocal effective model on the right. Circles denote sites, and lines denote couplings.}
    \label{fig: ISRNR}
\end{figure}

    A conventional and intuitive reason for the NHSE to appear is understood through the lens of nonreciprocity. An exemplary illustration of this phenomenon can be found in the Hatano-Nelson model \cite{Hatano1997VortexMechanics}. Alternatively, the NHSE can be induced by a combination of on-site gain/dissipation and complex couplings, but this mechanism may appear less intuitive \cite{Clerk2022IntroductionSchemes}.
    Consider the system depicted on the left-hand side of Fig.~\ref{fig: ISRNR}.
    It is a three-site non-Hermitian tight-binding Hamiltonian, with complex hopping parameters and one site featuring an imaginary on-site term.
    Specifically, if we enumerate the sites such that $u,v$ are the first two, the Hamiltonian is given by
    \begin{equation*}
        \ham = \begin{pmatrix}
            0 & 1 & e^{i \phi} \\
            1 & 0 & 1 \\
            e^{-i\phi} & 1 & i \alpha
        \end{pmatrix} \,.
    \end{equation*}
    Through an ISR to the red sites, the resulting effective model on the right-hand side is obtained. This reduced model exhibits a new on-site potential and hopping amplitudes, given by
    \begin{align*}
        A(E) &= \frac{1}{E-i\alpha},\\
        T_\pm(E) &= 1+\frac{e^{\pm i\phi}}{E-i\alpha},
    \end{align*}
    respectively. Notice how the hopping displays asymmetry in its magnitude, i.e., $|T_+(E)|\neq|T_-(E)|$, thereby indicating non-reciprocity within the model, in the same way as the Hatano-Nelson model. By employing an ISR, both avenues for realizing the NHSE---either directly by non-reciprocal couplings, or through reciprocal couplings but non-Hermitian on-site potentials---can be unified.
    
    In the following sections, we will modify our prototypical model slightly in order to have interesting, and emergent phenomena taking place. By considering a one-dimensional chain of fully connected four-site models instead of three-site ones, like in \cref{fig: ISRNR}, we are able to obtain an energy dependent skin-effect that induces localization on both sides of the system. 

        \begin{figure*}[!hbt]
    \centering
    \includegraphics[width=\textwidth]{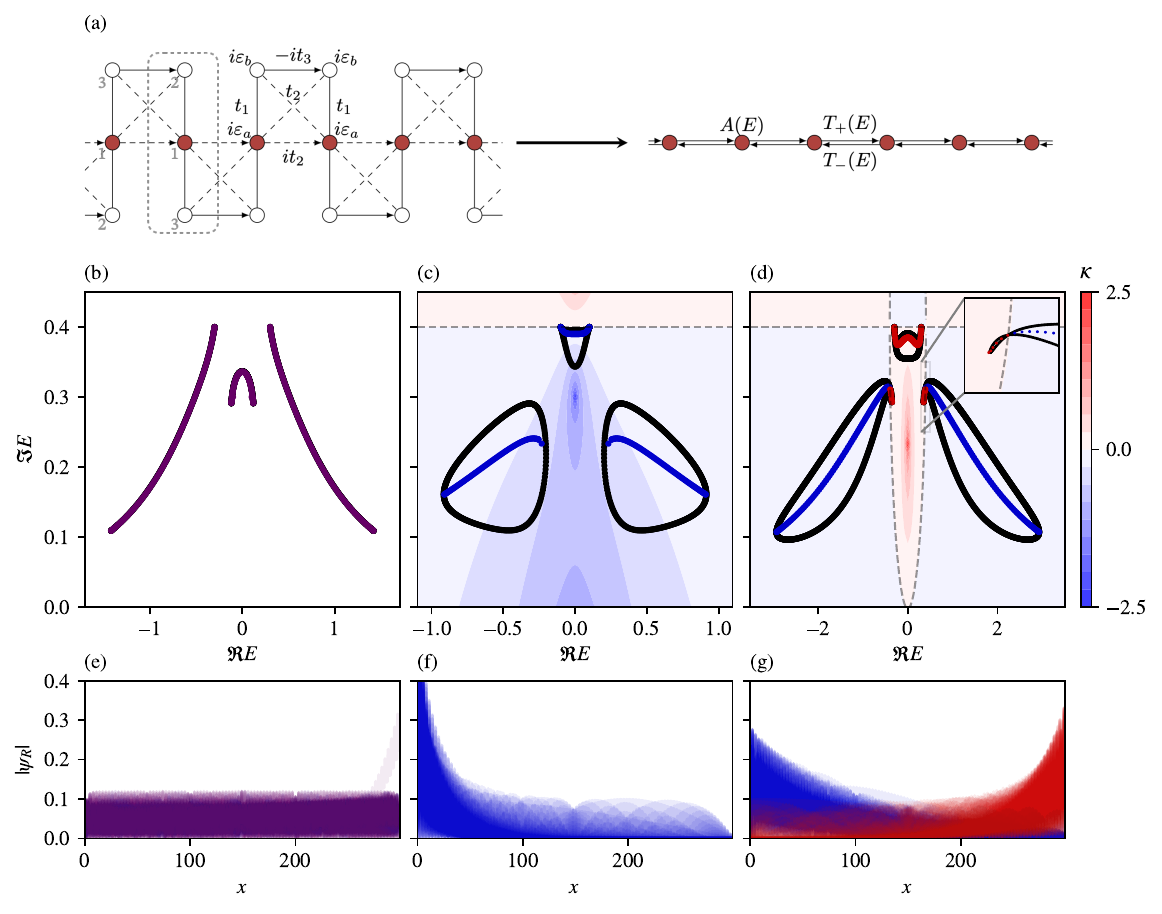}
    \caption{Emergent Hatano-Nelson model from ISR. (a) Lattice corresponding to the Bloch Hamiltonian given in \cref{eq:blochHN} and its ISR to the Hatano-Nelson model with energy dependent hopping and an energy dependent on-site term. The unit cell of the left chain is indicated by the gray dashed lines and the sites are labeled. (b-d) The complex energy spectra are presented for the system described in (a) using different parameter values: (b) $(\varepsilon_a, \varepsilon_b, t_1, t_2, t_3) = (0, 0.4,0,0.5,0.3)$; (c) $(\varepsilon_a, \varepsilon_b, t_1, t_2, t_3) = (0, 0.4, 0.4 , 0.2, 0.1)$; and (d) $(\varepsilon_a, \varepsilon_b, t_1, t_2, t_3) = (0, 0.4, 0.4 , 1, 0.3)$. The dots are color-coded, representing left localized (blue), right localized (red), or bulk-like (purple) modes, with black and colored lines indicating PBC and OBC, respectively. The background displays $\kappa(E)$, where larger absolute values of $\kappa$ indicate stronger localization of the corresponding skin mode. Dashed grey lines correspond to $\kappa(E) = 0$. (e-g) Right eigenstates are shown for the same parameter choices as the above panels, using the same coloring conventions.}
    \label{fig: HNSpectra}
\end{figure*}
   
    \section{Emergent Hatano-Nelson model} \label{Sec: Emergent HN model}
    Let us consider the model depicted in \cref{fig: HNSpectra}(a).
    Taking the unit cell as indicated in the figure, the Bloch Hamiltonian of this lattice is given by
    \begin{equation}
        H(k) = \begin{pmatrix}
            i\varepsilon_a-2t_2 \sin k & t_1 + t_2 e^{-ik} & t_2 + t_1 e^{-ik}\\
            t_1 + t_2e^{ik} & i \varepsilon_b & it_3\\
            t_2 + t_1 e^{ik} & -it_3 & i\varepsilon_b
        \end{pmatrix}.\label{eq:blochHN}
    \end{equation}
    Here $t_1$, $t_2$, and $t_3$ are real-valued hopping parameters, $\varepsilon_a$ and $\varepsilon_b$ are on-site gains or losses, and $k$ is the wave vector. Furthermore, we have set the lattice spacing to be equal to unity.
    
    We note at this point that the spectrum has a mirror symmetry with respect to the $\Re(E)=0$ line; cf. Figs.~\ref{fig: HNSpectra}(b) to \ref{fig: HNSpectra}(d).
    The mirror-symmetric spectrum stems from the fact that 
    \begin{equation}\label{Eq: spectral symmetry}
    	\begin{split}
    		\mathcal{S}H(k)\mathcal{S}^{-1}&=-H^*(-k), \\
    		\mathcal{S}&=(-\sigma_z)\oplus1,
    	\end{split}
    \end{equation}
    where $\sigma_z$ acts on the two sites of the unit cell.
    In the literature, this symmetry is better-known as $PHS^\dagger$ \cite{Kawabata2019PRX9041015SymmetryTopologyNonHermitianPhysics}, which is one of the two non-equivalent realizations of particle-hole symmetry in a non-Hermitian system. 
     
    If we simultaneously perform an ISR to all red sites of the full lattice, and take the Bloch-Hamiltonian of the resulting effective model, we obtain
\begin{equation}\label{Eq: HN Mom space}
    H_R(k,E) = A(E) + T_+(E) e^{ik} + T_-(E) e^{-ik},
\end{equation}
in which we recognize the Hatano-Nelson model with energy dependent on-site term $A(E)$, and hopping parameters $T_\pm(E) \equiv v(E) \pm g(E)$. Here
\begin{equation}\label{eq: paramHN}
\begin{split}
    A(E) &= i\left[\varepsilon_a+\frac{2\left(\varepsilon_b + iE\right) \left( t_1^2 + t_2^2 \right)}{\left(\varepsilon_b+iE\right)^2 +t_3^2}\right],  \\
    v(E) &= 2i\frac{\left(\varepsilon_b + iE\right)t_1 t_2}{\left(\varepsilon_b+iE\right)^2 +t_3^2}, \\
    g(E) &= i\frac{t_2t_3\left(t_3-t_2\right) + t_2 \left(\varepsilon_b + iE\right)^2 +t_1^2t_3}{\left(\varepsilon_b+iE\right)^2 +t_3^2}.
\end{split}
\end{equation} 
For the ordinary Hatano-Nelson model, i.e., no energy dependent parameters, it is well-known that the NHSE is present when $|T_+|\neq |T_-|$ \cite{Hatano1997VortexMechanics,Bergholtz2021ExceptionalSystems}. This condition still holds for our effective Hamiltonian. After some algebraic manipulations, it can be expressed as 
\begin{equation}\label{eq:vgvg}
    v_R(E)g_R(E) + v_I(E)g_I(E) \neq 0.
\end{equation}
Here, the subscripts $R$ and $I$ represent real and imaginary part, respectively. By substituting \cref{eq: paramHN} into \cref{eq:vgvg}, it follows that the NHSE is present when $t_1$, $t_2$ and $\varepsilon_b$ are all non-zero (see \cref{App: Calcs}). 

Let us now visualize the above statements in terms of the eigenvalues and eigenstates.
We start by inspecting a setup with $t_1 = 0$ and show its eigenvalue spectrum in \cref{fig: HNSpectra}(b).
The spectrum (denoted by a solid purple line) is the same for open boundary conditions (OBC) and periodic boundary conditions (PBC).
The (right) eigenvectors are depicted in \cref{fig: HNSpectra}(e) and show, as expected, no NHSE.
However, upon close inspection of \cref{fig: HNSpectra}(e), one can see a mode sitting at the right boundary.
This is a consequence of lattice termination and is elaborated upon in \cref{Sec:figure1Principle}.

Leaving the trivial case behind us, we next investigate a setup where $t_1$, $t_2$ and $\varepsilon_b$ are all non-zero. Specifically, we choose $(\varepsilon_a, \varepsilon_b, t_1, t_2, t_3) = (0, 0.4, 0.4 , 0.2, 0.1)$.
For this choice of parameters, the eigenvalue spectrum is shown in \cref{fig: HNSpectra}(c).
The black line represents the PBC eigenvalue spectrum, which forms three simple loops in the complex energy plane.
Importantly, the PBC spectrum now no longer coincides with the OBC spectrum (shown in blue).
The background color in this figure represents a contourplot of the skin length scale \cite{Bergholtz2021ExceptionalSystems}
\begin{equation} \label{eq: kappa}
	\kappa(E) \equiv \log\sqrt{\left|\frac{T_-(E)}{T_+(E)}\right|} \,.
\end{equation}
We note that $\kappa$ is energy-dependent, which is a  consequence of the energy-dependence of the system's hopping parameters.
This energy-dependence is an important difference to the ordinary Hatano-Nelson model. There, $\kappa$ is constant, such that all skin modes show the same length scale. 
In an emergent Hatano-Nelson model, on the other hand, each mode has its own skin length given by \cref{eq: kappa}.
In particular, a (right) PBC-eigenstate whose energy $E$ lies in a region with $\kappa < 0$ ($\kappa > 0$) will be localized at the system's left (right) boundary.
Now, since all of the system's OBC eigenvalues correspond to $\kappa < 0$, we expect all of the eigenstates to be left-localized.
This is indeed the case, as can be seen from \cref{fig: HNSpectra}(f).

Let us now modify the parameters to realize the so-called bipolar NHSE \cite{Song2019PRL123246801NonHermitianTopologicalInvariantsReal}. A system with a bipolar NHSE features two classes of right eigenstates: One being localized at the left boundary, and the other localized at the right boundary.
To find this phenomenon in our setup, we choose $(\varepsilon_a, \varepsilon_b, t_1, t_2, t_3) = (0, 0.4, 0.4 , 1, 0.3)$.
The system's eigenvalues are depicted in \cref{fig: HNSpectra}(d), which has an insect-like shape.
Again, PBC (black) and OBC (blue/red) spectra do not coincide, as expected from the fact that all $t_1$, $t_2$ and $\epsilon_b$ are non-vanishing.
What is interesting here is that $\kappa(E)$ can now take both positive and negative values.
In particular, we see that \cref{fig: HNSpectra}(d) splits into two regions: An outer, blue region, where $\kappa < 0$, and an inner, red region, where $\kappa > 0$.
These two regions are separated by the dashed-grey line that represents $\kappa(E) = 0$.
Now, since the OBC spectrum lies both in the inner and the outer region, our system features a bipolar NHSE: (Right) eigenstates whose energy $E$ lies in the blue region are left-localized, while eigenstates with $E$ lying in the red region are right-localized.
This is demonstrated in \cref{fig: HNSpectra}(g), where we show the system's right eigenstates, with blue/red color corresponding to the region in which the respective eigenvalue lies.

\begin{figure*}[!hbt]
    \centering
    \includegraphics[width=\textwidth]{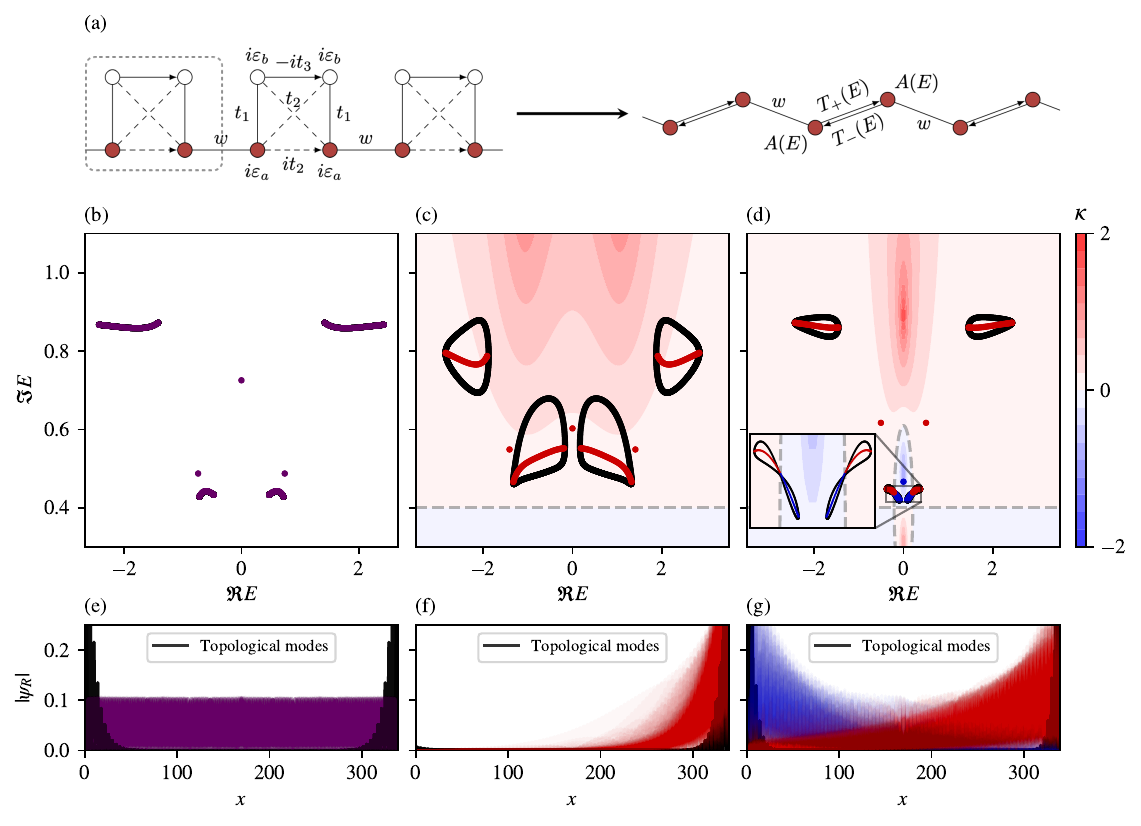}
    \caption{Emergent NH SSH model from ISR. (a) Lattice corresponding to the Bloch Hamiltonian given in \cref{Eq: Depleted Creutz Bloch} and its ISR to the NH SSH model with energy dependent hopping and an energy dependent on-site term. (b-d) The complex energy spectra are presented for the system described in (a) using different parameter values: (b) $(\varepsilon_a,\varepsilon_b,t_1,t_2,t_3)=(0.9, 0.4, 0, 0.5, 0.6)$; (c) $(\varepsilon_a,\varepsilon_b,t_1,t_2,t_3)=(0.9, 0.4, 1, 0.5, 0.9)$; and (d) $(\varepsilon_a,\varepsilon_b,t_1,t_2,t_3)=(0.9, 0.4, 0.2, 0.5, 0.2)$. In all figures we take $w=1.8$. The dots are color-coded, representing left localized (blue), right localized (red), or bulk-like (purple) modes, with black and colored lines indicating PBC and OBC, respectively. The background displays $\kappa(E)$, where larger absolute values of $\kappa$ indicate stronger localization of the corresponding skin mode. Dashed grey lines correspond to $\kappa(E) = 0$. (e-g) Right eigenstates are shown for the same parameter choices as the above panels, using the same coloring conventions. We also plot the isolated topological modes in black.}
    \label{fig: Creutz-Ladder}
\end{figure*}

The use of ISR does not limit itself to making predictions on the NHSE. On the contrary, it may also be used to explore topological properties of a given system. To illustrate this feature, we turn our attention to a different, but related model in the next section.

\section{Emergent non-Hermitian SSH Model} \label{Sec: SSH Reduc} 
    In this section, we study the system depicted in \cref{fig: Creutz-Ladder}(a), which is a modified version of the Creutz ladder \cite{Creutz1999EndFermions}. Each square forms a unit cell, interconnected by a real hopping parameter $w$, making this a four band model. The momentum space Hamiltonian for this system is given by 
\begin{equation}\label{Eq: Depleted Creutz Bloch}
	H(k)=\begin{pmatrix}
		i\varepsilon_a & t_1 & t_2 & it_2+we^{-ik} \\
	    t_1 & i\varepsilon_b & -it_3 & t_2 \\
		t_2 & it_3 & i\varepsilon_b & t_1 \\ 
		-it_2+we^{ik} & t_2 & t_1 & i\varepsilon_a
	\end{pmatrix},
\end{equation}
where all parameters are real-valued. The setup is similar to the one used by Lee \cite{Lee2016AnomalousLattice} to show the existence of an anomalous edge state. Our model was chosen such that its ISR to the red sites in \cref{fig: Creutz-Ladder}(a) results in an energy-dependent NH SSH model \cite{Su1979SolitonsPolyacetylene,Lieu2018TopologicalModel}, described by the following Bloch Hamiltonian: 
\begin{equation}\label{Eq: LAtent NH SSH Bloch}
	H_R(k,E)=\begin{pmatrix}
		A(E) & T_{+}(E)+we^{-ik} \\
		T_-(E)+we^{ik} & A(E)
	\end{pmatrix}. 
\end{equation}
Here
\begin{equation}\label{Eq: Parameters NHSSH}
    \begin{split}
        A(E)&=i\left[\varepsilon_a+\frac{\left(\varepsilon_b + iE\right) \left( t_1^2 + t_2^2 \right)}{\left(\varepsilon_b+iE\right)^2 +t_3^2}\right]
    \end{split}
\end{equation}
and $T_\pm(E)$ are the same as given in \cref{Eq: HN Mom space,eq: paramHN}.
This reduction is graphically depicted in \cref{fig: Creutz-Ladder}(a). We observe that our model features a rich variety of phases, from the NHSE to topological edge modes. We note that this model enjoys the same $PHS^\dagger$ symmetry as the previous three-band model \cref{Eq: spectral symmetry}. However $\mathcal{S}$ now must be built from a different partitioning and is given by $\mathcal{S}=(-\sigma_z)\oplus\mathbbm{1}_{2\times 2}$. 

\subsection{Onset of the NHSE}
Similar to the Hatano-Nelson model, the NHSE is present in the NH SSH model whenever $|T_{+}|\neq|T_{-}|$. 
By analogy, for our emergent NH SSH model, this results in the constraint equation $v_R(E)g_R(E)+v_I(E)g_I(E) \neq 0$. In terms of the model parameters, this leads to the condition that the NHSE is present when $t_1$, $t_2$, and $\varepsilon_b$ are not equal to zero. This is illustrated in \cref{fig: Creutz-Ladder}, where the three possible scenarios are depicted. First, \cref{fig: Creutz-Ladder}(b) shows the case without skin effect, clearly indicated by similar band structures for OBC and PBC, and the corresponding right eigenstate in \cref{fig: Creutz-Ladder}(e).  \cref{fig: Creutz-Ladder}(c) shows the band structure when the skin-effect is present, but only in one direction, as indicated by $\kappa(E)>0$. The modes localize on the right-hand side, as shown in \cref{fig: Creutz-Ladder}(f). Finally, \cref{fig: Creutz-Ladder}(d) shows the band structure when the bipolar skin-effect is present, which can be understood from the contourplot of $\kappa(E)$, showing both regions of $\kappa>0$ (red) and $\kappa<0$ in blue. In all three situations, one can observe the presence of six topological edge modes, coming in three pairs of two degenerate modes pinned at the same energy. These are shown in black in Figs.~\ref{fig: Creutz-Ladder}(e)-\ref{fig: Creutz-Ladder}(g). We will now investigate the properties of these topological modes.
\begin{figure}[!t]
    \centering
    \includegraphics[width=\columnwidth]{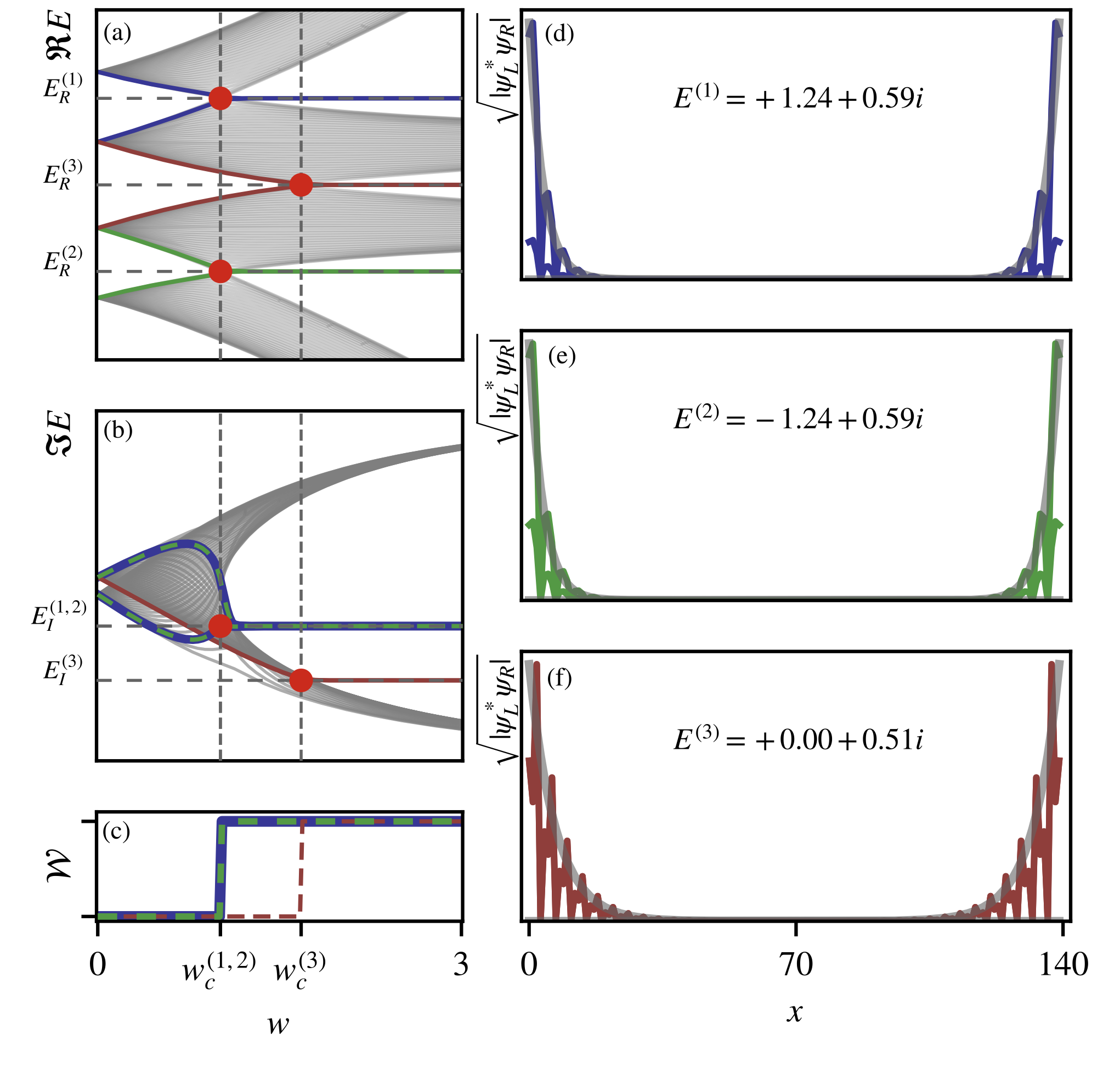}
    \caption{Topological phase transitions of the emergent NH SSH model as a function of $w$. (a) Real and (b) imaginary part of the energy spectrum, for an open chain consisting of $N_c=35$ unit cells. The spectra are taken for the parameter choice $(\varepsilon_a,\varepsilon_b,t_1,t_2,t_3)=(0.9, 0.4, 1, 0.5, 0.6)$.(c) The winding number given by \cref{Eq: Winding number def}, calculated at the three special energies $E_t$, clearly shows its quantization and the critical points $w$. (d-f) Corresponding topological edge modes, at $w = 3$, plotted together with the calculated exponential envelope in gray, with penetration depth given by \cref{eq: xi}. The three different colors blue, green, red are used consistently to mark the different edge modes [(d) to (f)], the behavior of their energies [(a) and (b)], and the values of the corresponding winding number (c).}
    \label{fig: Topo Phase Transitions}
\end{figure}
\subsection{Topological Edge Modes} 
Interestingly, we can also predict the existence of topological edge modes in the four-band model using the reduced NH SSH chain. The winding number that determines the topological phase transition for a sublattice-symmetric 1D Hamiltonian (of which the NH SSH is an example) is given by \cite{Kawabata2019PRX9041015SymmetryTopologyNonHermitianPhysics} 
\begin{equation}\label{Eq: Winding number def}
    \mathcal{W}=\int_{-\pi}^\pi\frac{dk}{4\pi i}\Tr\left[\sigma_zH^{-1}(k)\frac{dH(k)}{dk}\right].
\end{equation}
In our case, this expression becomes energy dependent and is only applicable when 
\begin{equation}\label{Eq: condition Top Edge Mode}
    A(E)=E,
\end{equation}
where $A(E)$ is defined in \cref{Eq: Parameters NHSSH}. This is because \cref{Eq: Winding number def} is only well-defined for sublattice-symmetric systems, which in our case is a latent symmetry appearing at energies satisfying \cref{Eq: condition Top Edge Mode}. This means that we must consider a Hamiltonian $\Tilde{H}(k)\equiv H(k,E_t)-E_t\mathbbm{1}_{2\times2}$, where $E_t$ is a solution of \cref{Eq: condition Top Edge Mode}. For every energy satisfying this constraint, in the topological phase, there is a degenerate pair of edge states pinned at that energy. In fact, the pair is quasi-degenerate, as a consequence of the finite size of the lattice. For the model at hand, there are three energies at which the transition takes place because \cref{Eq: condition Top Edge Mode} has three solutions. Explicit calculations of this winding number (see \cref{App: Calcs} for an analytic derivation) lead to
\begin{equation}\label{Eq: w transition}
\mathcal{W}(E_t)=\left\{\begin{alignedat}{5}
    & 0, \ \ \ \text{if } |w|<\sqrt{|v^2(E_t)-g^2(E_t)|}  \\
               & 1, \ \ \ \text{if } |w|>\sqrt{|v^2(E_t)-g^2(E_t)|}
\end{alignedat}\right. .
\end{equation}
Substituting the solutions $E_t$ into \cref{Eq: w transition} yields the three critical values $w_c$ at which pairs of topological edge modes appear. Figs.~\ref{fig: Topo Phase Transitions}(a) and \ref{fig: Topo Phase Transitions}(b) show the real and imaginary parts of the energy spectrum, respectively. The horizontal dashed lines show the calculated absolute values of the complex transition energies $E^{(j)}_t=E^{(j)}_R+E^{(j)}_I$, while the vertical dashed lines indicate the value of the predicted critical hopping parameter $w_c$. Robust boundary modes that persist beyond the transition point are visible in red, green and blue. The presence of these modes can be quantified by calculating the winding number given by \cref{Eq: w transition}, as shown in Fig.~\ref{fig: Topo Phase Transitions}(c). There is a clear jump to $\mathcal{W}(E_t)=1$ when the critical hopping $w_c$ is reached. Figs.~\ref{fig: Topo Phase Transitions}(d)-\ref{fig: Topo Phase Transitions}(f) show the corresponding edge states, with the same color coding, at $w=3$. The values of the calculated transition energies $E_t$ are indicated in the middle of each figure. Notice that to properly visualize the edge modes in the presence of the NHSE, $\sqrt{|\psi_L^*\psi_R|}$ is plotted rather than $|\psi_R|$. Moreover, it is important to highlight that, for each transition energy, the emergent sublattice symmetry of the reduced model ensures that the edge modes appear in pairs. This is visible in Figs.~\ref{fig: Topo Phase Transitions}(d)-\ref{fig: Topo Phase Transitions}(f), where a pair of edge states is shown for each energy $E_t$. For a further investigation of the line-gap closings of the emergent NH SSH model, we refer the reader to \cref{app: TPT}. As a final note, we see that the penetration depth of these edge states is also energy dependent, and is given by \cite{Bergholtz2021ExceptionalSystems,Kunst2018PRL121026808BiorthogonalBulkBoundaryCorrespondenceNonHermitian} 
\begin{equation}
    \begin{split}
        \xi_L(E)&=\frac{1}{\log\left|\frac{v(E)-g(E)}{w}\right|}, \\
        \xi_R(E)&=\frac{1}{\log\left|\frac{v(E)+g(E)}{w}\right|},
    \end{split}
\end{equation}
where the subscript $L$ ($R$) stands for left (right) eigenvectors. This explains the different localization lengths observed in Figs.~\ref{fig: Topo Phase Transitions}(d)-\ref{fig: Topo Phase Transitions}(f). In the biorthogonal formulation the exponential envelope follows $\exp{-x_j/2\xi_{LR}}$, where $x_j=ja$ and
\begin{equation}\label{eq: xi}
    \xi_{LR}=\frac{\xi_L+\xi_R}{\xi_R\xi_L}.
\end{equation}
This envelope is plotted in grey alongside the edge states in Figs.~\ref{fig: Topo Phase Transitions}(d)-\ref{fig: Topo Phase Transitions}(f).

\begin{figure}[!t]
    \centering
    \includegraphics[width=0.9\linewidth]{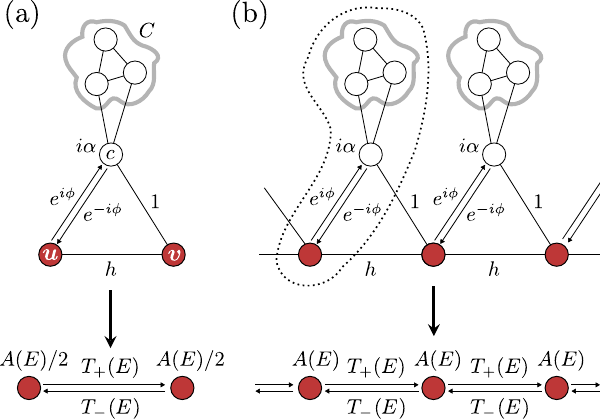}
    \caption{(a) ISR of the lossy, complex hopping model above onto the red sites yields the non-reciprocal effective model below. (b) Lattice realization of (a). The unit cell is marked by a dashed line.}
    \label{fig:ISRNR_Extension}
    \end{figure}
    \section{Generalized construction principles} \label{Sec: Construction Principles}
    
    The models treated in the previous sections are individual examples of setups whose ISR has the form of an effective Hatano-Nelson or NH SSH model. In the following, we will show that one can systematically construct large families of such systems. The procedure will always be the same: an individual unit cell is built, such that its ISR to two specific sites yields equal on-site potentials, and nonreciprocal hoppings between them. Subsequently, these unit cells are connected such that either (i) a Hatano-Nelson, or (ii) a NH SSH model is obtained.
    
\begin{figure}[!b]
    \centering
    \includegraphics[width=0.9\linewidth]{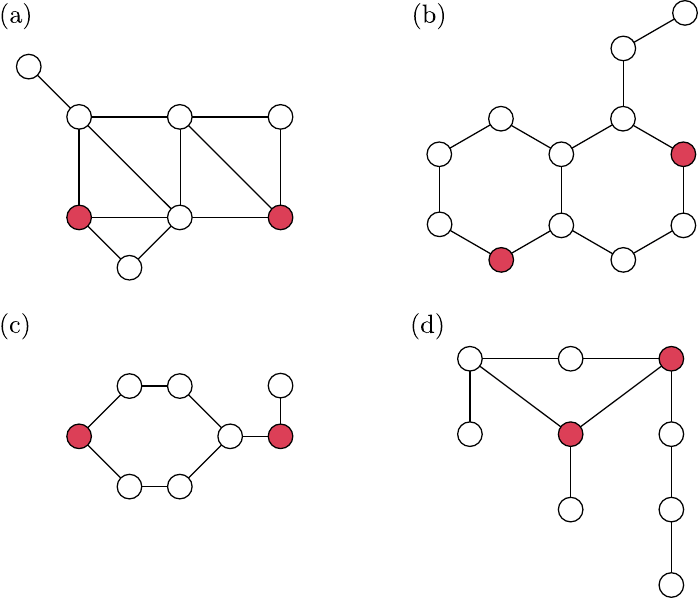}
    \caption{Different systems with latently symmetric sites (marked in red). In all four sytems, each line corresponds to a coupling of strength one (see Refs. \cite{Rontgen2020PRA101042304DesigningPrettyGoodState,Morfonios2021LAaiA62453CospectralityPreservingGraphModifications} for more details regarding the design of latently symmetric setups).}
    \label{fig:latentSymmetry}
    \end{figure}
    
    \subsection{Emergent Hatano-Nelson models} \label{Sec:figure1Principle}

\begin{figure*}[!hbt]
    \centering
    \includegraphics[width=0.85\linewidth]{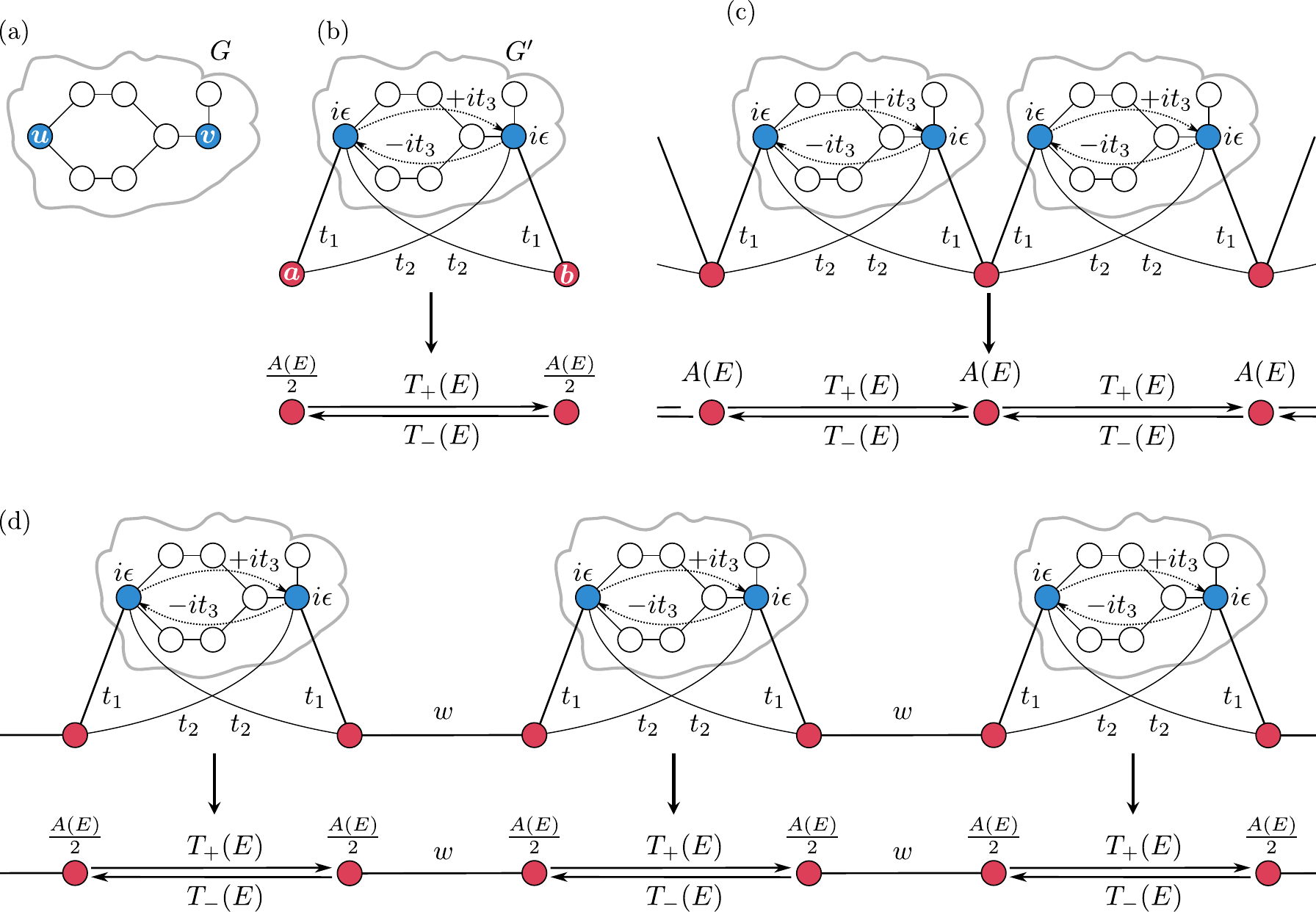}
    \caption{Construction scheme for the emergent Hatano-Nelson and NH SSH models. (a) The starting point: A simple setup with latently symmetric sites $u,v$. (b-d) ISR of the lossy, complex hopping model above onto the red sites yields the non-reciprocal effective model below.}
    \label{fig:HNspectrum_Extension}
\end{figure*}

	\subsubsection{Construction principle A} \label{sec:figure1Principle}
    The starting point for the first construction principle is a finite structure that has a non-reciprocal ISR. The graphical representation of the model is sketched in \cref{fig:ISRNR_Extension}(a). The system consists of two lower sites $u$ and $v$ (marked in red), which are connected to a third site $c$ via Hermitian couplings $\exp(i \phi)$ and $1$.
    The site $c$ has complex on-site potential $i \alpha$ and is further coupled to a (possibly very large) network $C$.
    For simplicity, we demand that the couplings in the network $C$, as well as the couplings between this network and the site $c$ are real-valued. However, the sites in $C$ could have complex on-site potentials.
    Denoting the Hamiltonian of the resulting total system by $\ham$, its ISR to the two sites $u$ and $v$ yields (see \cref{app:constructionPrinciples})
    \begin{equation*}
    \mathcal{R}_S(E,H) = 
    \begin{pmatrix}
        A(E)/2 & T_{+}(E) \\
        T_{-}(E) & A(E)/2
    \end{pmatrix} \,.
    \end{equation*}
    Note that the exact form of $A(E)$ depends on the details of the network $C$.
    
    This finite building block is now used to construct a lattice, as shown in \cref{fig:ISRNR_Extension}(b), with the unit cell comprised of one site $c$, one network $C$, and \emph{one of the two} red sites.
    Applying the ISR to all red sites, a Hatano-Nelson model emerges, as depicted in the lower part of \cref{fig:ISRNR_Extension}(b).
    Importantly, since each red site of the lattice is coupled to two clouds, its on-site potential after the ISR now reads $2 \cdot A(E)/2 = A(E)$. In an open chain, the sites on the left and right end, however, will have an on-site potential of $A(E)/2$ \footnote{We note that this is the reason why the non-topological edge states, as seen in Fig.~\ref{fig: HNSpectra}(e), appear.}.

    \subsubsection{Construction principle B}
    \label{sec:figure2Principle}
    The second construction principle of emergent Hatano-Nelson models relies on the concept of latent symmetry \cite{Smith2019PA514855HiddenSymmetriesRealTheoretical}. Given a system $G$, two sites $S=\{u,v\}$ are latently reflection symmetric if the ISR over them has the form 
        \begin{equation} \label{eq:latentSymmetry}
    \mathcal{R}_S(E,G) = 
    \begin{pmatrix}
        \mathcal{A}(E) & \mathcal{B}(E) \\
        \mathcal{B}(E) & \mathcal{A}(E)
    \end{pmatrix} \,,
    \end{equation}
    that is, if $\mathcal{R}_S(E,H)$ commutes with the permutation matrix $P := \sigma_x$. In \cref{fig:latentSymmetry}, a number of setups with latently symmetric sites (marked in red) are shown. A broader overview over this topic is given in Ref. \cite{Rontgen2022LatentSymmetriesIntroduction}.

\begin{figure*}[!hbt]
        \centering
        \includegraphics[width=\textwidth]{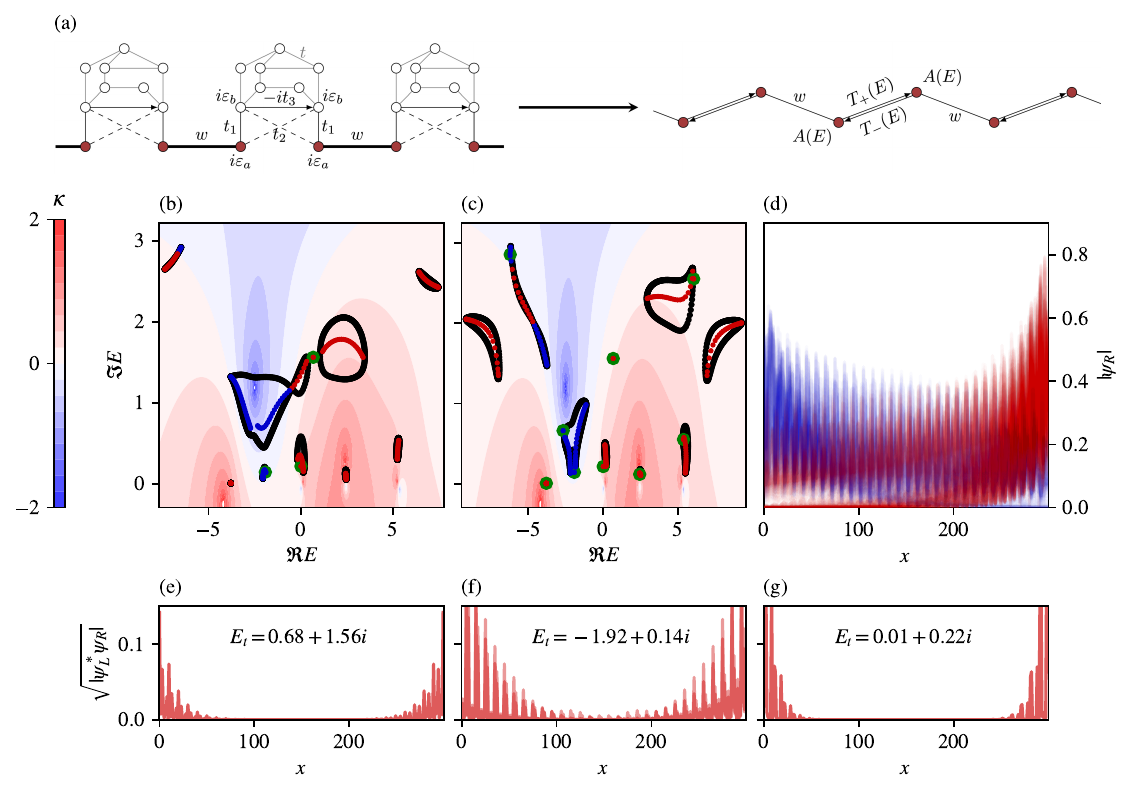}
        \caption{Example of a generalized construction, with a network of eight sites connected to the two sites to which the ISR is applied. (a) The model and its ISR. (b) PBC (black) and OBC (right NHSE in red, left NHSE in blue) spectra of a system with the following parameters  $(a,b,t_1,t_2,t_3,t,w)=(1.19, 3.74,1.18,2.97,4.24,2.03,2.5)$, together with the three predicted edge mode energies (with double degeneracy) shown with green circles at this particular $w$ value. Once again, the contour plot shows the skin depth $\kappa(E)$, with its values shown on the color bar in the left. (c) Same as (b), but with $w=6.5$, where we now see all possible edge modes appear, at 9 different energies, giving a total of 18 edge modes. (d) Right-eigenstate amplitudes corresponding to the parameter choice given in (a). (e-g) Amplitudes of all six eigenstates (two in each figure) that appear in (a), and their corresponding energies, shown in the biorthogonal basis.}
        \label{fig: 10-site NHSSH example}
    \end{figure*}     

    The construction scheme is sketched in Figs.~\ref{fig:HNspectrum_Extension}(a) to \ref{fig:HNspectrum_Extension}(c).
    \cref{fig:HNspectrum_Extension}(a) depicts a real-symmetric subsystem $G$ (marked by a cloud) in which two sites $u$ and $v$ (marked in blue) are latently symmetric.
    In other words, if one would perform the ISR over $u,v$, one would obtain \cref{eq:latentSymmetry}.
    The key here is that the latent symmetry guarantees the existence of a matrix $Q$ that commutes with the subsystem, i.e. $Q G = G Q$, and which (i) permutes the sites $u$ and $v$, (ii) is block-diagonal, and (iii) fulfills $Q^{-1} = Q^T = Q$ \cite{Rontgen2021PRL126180601LatentSymmetryInducedDegeneracies}. 
    In the following, we shall use this matrix extensively.
    
    In the next step, this subsystem is modified by adding complex on-site potentials $i \epsilon$ to $u$ and $v$, which are then connected via Hermitian hoppings $i t_3$. 
    Note that this breaks the latent symmetry: If we denote the resulting modified subsystem by $G'$, then the isospectral reduction of $G'$ over $u,v$ would read
            \begin{equation*}
    	\mathcal{R}_S(E,G') = 
    	\begin{pmatrix}
    		\mathcal{A}(E) + i \epsilon& \mathcal{B}(E) + i t_{3} \\
    		\mathcal{B}(E) - i t_{3}& \mathcal{A}(E) + i \epsilon
    	\end{pmatrix} \,,
    \end{equation*}
    which does not commute with $P$.
    However, we have $\mathcal{R}_S(E,H)\,  P = P\, \mathcal{R}_S(E,H)^T$.
    Due to the favourable properties of $Q$---in particular, its block-diagonal form---, it can be shown that $Q G' = G'^{T} Q$.

    At this point, two additional sites $a$ and $b$ (marked in red) are coupled to the subsystem $G'$ with hoppings $t_{1}$ and $t_{2}$. 
    Again employing the favourable properties of $Q$, it can be easily shown that the Hamiltonian $H$, describing the total system depicted in \cref{fig:HNspectrum_Extension}(b), obeys $Q' H = H^{T} Q'$.
    Here, the matrix $Q' = P \oplus Q$, with the permutation matrix $P$ acting on the two red sites $a$ and $b$.
	Analogously, it can be shown that the ISR has the form
      \begin{equation*}
    	\mathcal{R}_S(E,H) = 
    	\begin{pmatrix}
    		A(E)/2 & T_{+}(E) \\
    		T_{-}(E) & A(E)/2
    	\end{pmatrix} \,.
    \end{equation*}
    Note that one can relate $\mathcal{A}(E), \mathcal{B}(E)$ to $A(E), T_{\pm}(E)$, though we omit the exact relation here.
    
    Again, a lattice can be built by taking one red site and one subsystem $G'$ as a unit cell, see \cref{fig:HNspectrum_Extension}(c). Taking the ISR to all red sites of this lattice doubles the on-site potential, which then becomes $A(E)$ instead of $A(E)/2$.

\subsection{Emergent NH SSH model}
    In the previous \cref{sec:figure1Principle},lattices were built by taking one red site and one subsystem $G'$ as a unit cell, which resulted in emergent Hatano-Nelson models.
    One could, however, also build a lattice by taking one subsystem $G'$ and  \emph{two} red sites as a unit cell, and then connect neighbouring unit cells via an additional coupling $w$, as shown in the upper part of \cref{fig:HNspectrum_Extension}(d). 
    This results in an emergent NH SSH model, which is depicted in the lower part of \cref{fig:HNspectrum_Extension}(d).
    Note that, by removing all complex couplings and on-site potentials, one would obtain an effective version of the conventional SSH model.
    Such emergent SSH models have been very recently investigated in Ref.~\cite{Rontgen2023LatentSuSchriefferHeegerModels}.
    
    Before concluding this work, we investigate a specific realization of the above procedure that results in an emergent NH SSH model.
    The setup and its ISR are depicted in  \cref{fig: 10-site NHSSH example}(a). The resulting OBC (red and blue) and PBC (black) spectra are shown in \cref{fig: 10-site NHSSH example}(b) and \ref{fig: 10-site NHSSH example}(c), where the intercell hopping parameter is $w=2.5$ in (b) and $w=6.5$ in (c). This leads to the appearance of six topological edge states in (b), and eighteen in (c) (two doubly degenerate modes per energy). The overlaid transparent green circles indicate the presence of these edge modes in the OBC spectrum. The right eigenstates corresponding to the parameter choice in (b) are shown in \cref{fig: 10-site NHSSH example}(d). There, one can again observe the energy-dependent skin effect. Figs.~\ref{fig: 10-site NHSSH example}(e)-\ref{fig: 10-site NHSSH example}(g) show all six edge states that exist in (b), and their corresponding energies. Note that, since there are nine solutions to the equation $A(E)-E=0$, the total amount of possible edge states is eighteen. The double degeneracy of each energy solution is, once again, a result of the emergent sublattice symmetry.

	\section{Conclusion} \label{Sec: Conclusion}
    Across many branches of physics, toy models are an essential tool to understand the key features of a given theory.
    In non-Hermitian physics, two such models are the Hatano-Nelson and the non-Hermitian Su-Schrieffer-Heeger (NH SSH) model.
    Despite their simple structure---their unit cells comprise only a single or two sites, respectively---these one-dimensional models host non-trivial boundary phenomena.
    In this work, we have used recent graph-theoretical insights to design systems whose so-called isospectral reduction---akin to an effective Hamiltonian---takes the form of either of these models.
    This procedure keeps the structure of the toy model, while simultaneously enriching it by making the couplings and on-site potentials energy-dependent.
    Specifically, we have shown that this leads to \emph{emergent Hatano-Nelson} or  \emph{emergent NH SSH models} featuring a two-sided non-Hermitian skin effect, caused by an energy dependence of the skin localization length $\kappa(E)$.
    This energy-dependence allows different states to be localized on different ends of the system, and to have different localization strengths.
    For the emergent NH SSH models, we observed topological edge modes which we could further link to a quantization of the winding number.
    In all cases, the original system---whose isospectral reduction becomes a Hatano-Nelson or NH SSH model---features only reciprocal (though complex-valued) couplings, with non-Hermiticity entering through complex on-site potentials (gain/loss).

    We emphasize that the methods and ideas presented in this work are not limited to one-dimensional non-Hermitian Hamiltonians, but are rather generic. For example, they can be applied to higher-dimensional setups, which reduce under the isospectral reduction to paradigmatic models.  An interesting avenue to explore would be the realization of these construction principles in different platforms, such as photonic or acoustic waveguides, electric circuits, or mechanical metamaterials.

	\begin{acknowledgments}
    A.M. and C.M.S. acknowledge the TOPCORE project with project number OCENW.GROOT.2019.048 which is financed by the Dutch Research Council (NWO). L.E. and C.M.S. acknowledge the research program “Materials for the Quantum Age” (QuMat) for financial support. This program (registration number 024.005.006) is part of the Gravitation program financed by the Dutch Ministry of Education, Culture and Science (OCW). 
    V.A. is supported by the EU H2020 ERC StG ”NASA” Grant Agreement No. 101077954. 
    L.E. and A.M. contributed equally to this work.
	\end{acknowledgments}
	
    \bibliography{BibMalte, BibAnouar} 

\appendix
\begin{widetext}
\section{Detailed calculations for the NHSE and topology}\label{App: Calcs}
\subsection{Constraints on the NHSE}
As stated in the main text, the NHSE appears whenever $|T_+(E)|\neq |T_{-}(E)|$, where $T_{\pm}(E)$ is given by \cref{eq: paramHN}. From \cref{Eq: HN Mom space,eq: paramHN}, we derive the following constraint equation
\begin{equation}\label{Eq: No NHSE constraint}
        \frac{2t_1t_2(\varepsilon_b-E_I)\left[ t_1^2t_3+t_2 \left[ -t_2 t_3 + t_3^2 + (\varepsilon_b-E_I)^2 + E_R^2 \right] \right]}{\left[ (\varepsilon_b-E_I)^2 + (t_3 - E_R)^2 \right]\left[ (\varepsilon_b-E_I)^2 + (t_3 + E_R)^2 \right]}=0.
\end{equation}
From this constraint equation, we can easily see that the NHSE disappears whenever $t_1=0$, $t_2=0$ or $\varepsilon_b=0$. \cref{Eq: No NHSE constraint} also allows us to determine the contour curves of $\kappa(E)=0$, which were plotted in \cref{fig: HNSpectra}(d) and \cref{fig: Creutz-Ladder}(d). These are given by 
\begin{equation}
    \begin{split}
        E_I&=\varepsilon_b, \\
        E_I&=\varepsilon_b\pm\frac{\sqrt{-t_2 \left[E_R^2 t_2 + t_3 \left(t_1^2 + -t_2^2 + t_2t_3\right)\right]}}{t_2}.
    \end{split}
\end{equation}

\subsection{Determining the topological transition energies of the four-band model} 
The constraint equation that allows us to find the topological phase transition energies is $A(E)-E=0$. This condition yields the cubic equation
\begin{equation}\label{Eq: root eqs SSH transition}
    E^3 -i(\varepsilon_a+2\varepsilon_b)E^2 -\left( t_1^2 + t_2^2 + t_3^2 + 2\varepsilon_a\varepsilon_b + \varepsilon_b^2 \right)E + i\left[ \varepsilon_at_3^2 + \varepsilon_b\left(t_1^2+t_2^2\right)+ \varepsilon_a\varepsilon_b^2\right] = 0.
\end{equation}
This equation can be solved exactly and the solutions are used to determine the transition energies. These are given by 
\begin{equation}
    \begin{split}
    E_1&=-\frac{1}{3}\left[A-2^{\frac{1}{3}}\mathcal{F}_1-2^{-\frac{1}{3}}\mathcal{F}_2\right], \\
    E_2&=-\frac{1}{3}\left[A+(-2)^\frac{1}{3}\mathcal{F}_1-\frac{1}{2}(-2)^\frac{2}{3}\mathcal{F}_2\right], \\
    E_3&=-\frac{1}{3}\left[A+(-1)^\frac{2}{3}2^\frac{1}{3}\mathcal{F}_1-(-2)^{-\frac{1}{3}}\mathcal{F}_2\right],
    \end{split}
\end{equation}
where we have defined the following expressions
\begin{align*}
    \mathcal{F}_1&=\frac{A^2-3B}{\left[-2A^3+9AB-18C+3\sqrt{3(4B-A^2)B^2+6A(2A^2-9B)C+81C^2}\right]^\frac{1}{3}}, \\
    \mathcal{F}_2&=\left[-2A^3+9AB-18C+3\sqrt{3(4B-A^2)B^2+6A(2A^2-9B)C+81C^2}\right]^\frac{1}{3},
\end{align*}
with
\begin{align*}
    A&=-i\varepsilon_a+2i\varepsilon_b, \\
    B&=-t_1^2-t_2^2-t_3^2-2\varepsilon_a\varepsilon_b-\varepsilon_b^2, \\
    C&=i\varepsilon_at_3^2+i\varepsilon_b\left(t_1^2+t_2^2\right)+i\varepsilon_a\varepsilon_b^2.
\end{align*}
\newpage
\subsection{Quantization of the winding number} The winding number for the reduced model in \cref{Sec: SSH Reduc} is given by \cref{Eq: Winding number def}. Working out the trace of the product of matrices yields the following equation

\begin{equation*}
    \mathcal{W}(E_t)=\int_{-\pi}^\pi\frac{dk}{2\pi i}\frac{i e^{ik} w \left(w + 
   a(E_t) \cos k\right)}{(e^{ik} a(E_t) + 
   w) (a(E_t) + e^{ik} w)},
\end{equation*} 
where $a(E_t)=\sqrt{v^2(E_t)-g^2(E_t)}$. We can turn this into a contour integral on the unit circle $S^1$ by letting $z=e^{ik}$. This gives
\begin{align*}
    \mathcal{W}(E_t)&=\oint_{S^1}\frac{dz}{2\pi i}\frac{1}{2}\frac{w\left[2w+a\left(z+\frac{1}{z}\right)\right]}{\left(za+w\right)\left(a+zw\right)}, \\
    &=\oint_{S^1}\frac{dz}{2\pi i}\frac{1}{2a}\frac{2wz+a\left(z^2+1\right)}{z\left(z+\frac{a}{w}\right)\left(z+\frac{w}{a}\right)}, \\
    &\equiv\oint_{S^1}\frac{dz}{2\pi i}f(z),
\end{align*}
where we omitted the argument $E_t$ in $a$, for brevity. The function $f(z)$ has two poles inside the unit circle and one outside. The pole $z_0=0$ is always inside, while $z_1=w/a$ is inside if $|w|<|a|$ (making $z_2=a/w$ outside). Using the residue theorem, we then have the following conditions
\begin{equation*}
\mathcal{W}(E_t)=\left\{\begin{alignedat}{5}
    & \text{Res}(f,z_0)+\text{Res}(f,z_1), \ \ \ \text{if } |w|<|a(E_t)|  \\
               & \text{Res}(f,z_0)+\text{Res}(f,z_2), \ \ \ \text{if } |w|>|a(E_t)|
\end{alignedat}\right.
\end{equation*}
Since the poles are of order 1, the residues are simply given by 
\begin{align*}
    \text{Res}(f,z_0)&=\lim_{z\to z_0}(z-z_0)f(z)=\frac{1}{2}, \\
    \text{Res}(f,z_1)&=\lim_{z\to z_1}(z-z_1)f(z)=-\frac{1}{2}, \\
    \text{Res}(f,z_2)&=\lim_{z\to z_2}(z-z_2)f(z)=\frac{1}{2},
\end{align*}
which finishes the proof of the quantization of the winding number,
\begin{equation*}
\mathcal{W}(E_t)=\left\{\begin{alignedat}{5}
    & 0, \ \ \ \text{if } |w|<\sqrt{|v^2(E_t)-g^2(E_t)|}  \\
               & 1, \ \ \ \text{if } |w|>\sqrt{|v^2(E_t)-g^2(E_t)|}
\end{alignedat}\right. .
\end{equation*}
\newpage

\section{Topological Phase Transitions in the Emergent NH SSH Model}\label{app: TPT}
\begin{figure*}[!b]
    \centering
    \includegraphics[width=\textwidth]{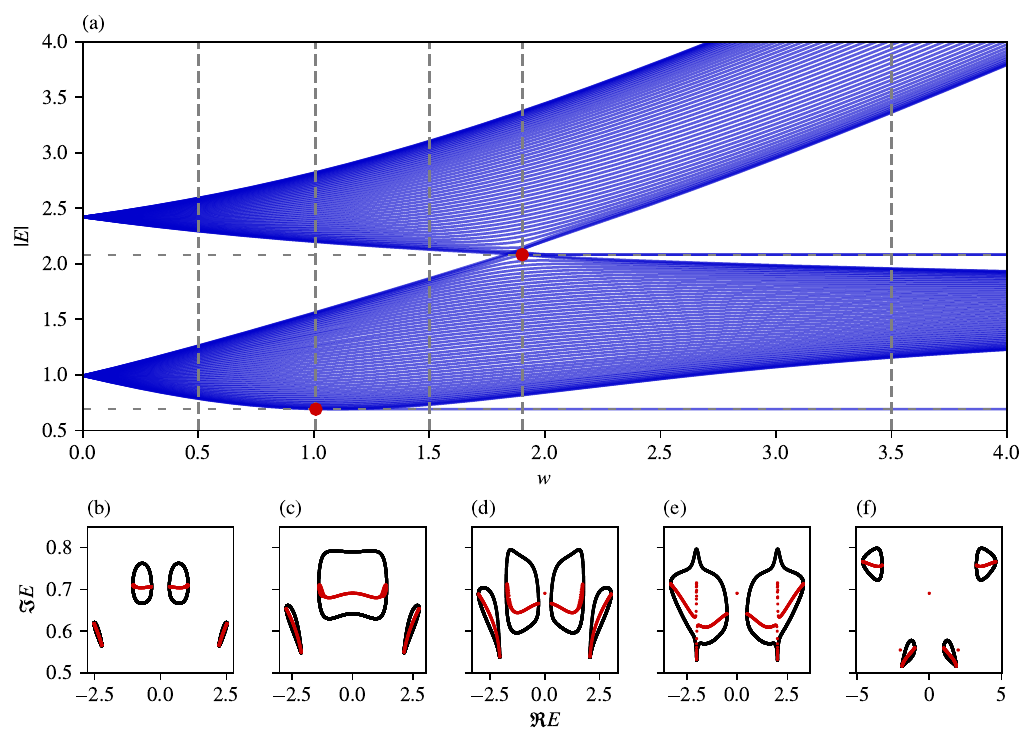}
    \caption{Behavior of the band structure (absolute value) of the emergent NH SSH model as described in \cref{Sec: SSH Reduc}. (a) Absolute value of the spectrum for $(\varepsilon_a,\varepsilon_b,t_1,t_2,t_3)=(0.8, 0.5, 1, 0.7, 1.6)$ as a function of $w$, for OBC with $N=50$ unit cells. 
    There are three topologically distinct phases. The transitions are denoted by a red dot, sitting at the intersection between the edge modes' energies (horizontal dashed gray line) and the critical $w$ values. (b-f) The spectrum in the complex plane for fixed $w$ corresponding to the vertical dashed gray lines in (a). Red and black points denote the OBC and PBC spectra, respectively. (b) Trivial phase, (c) first line-gap closing, (d) topological phase with one (degenerate) pair of edge modes, (e) second line-gap closing, (f) topological phase with three (degenerate) pairs of edge modes.}
    \label{fig: TPT}
\end{figure*}
In this section, we analyze the different topological phases of the model described by \cref{Eq: Depleted Creutz Bloch}, which reduces to the NH SSH model. \cref{fig: TPT}(a) shows the absolute value of the spectrum as a function of $w$ for OBC. Here, one can observe the emergence of edge modes for different values of $w$. The topological nature of these modes is further elaborated upon in the main text. We then take `slices' of \cref{fig: TPT}(a) at fixed $w$ to obtain Figs.~\ref{fig: TPT}(b)-\ref{fig: TPT}(f), where both OBC (red) and PBC (black) spectra are shown in the complex plane. The chosen values of $w$ correspond to the vertical dashed lines in \cref{fig: TPT}(a). Figs.~\ref{fig: TPT}(b), \ref{fig: TPT}(d), and \ref{fig: TPT}(f) show a trivial phase and two distinct topological phases, respectively. Figs.~\ref{fig: TPT}(c) and \ref{fig: TPT}(e) show the two (OBC) line-gap closings. It should be noted that, due to the breakdown of the bulk boundary correspondence for non-Hermitian systems, the line-gap closings for OBC and PBC do not take place at the same point in parameter space.

\section{Additional information for construction principles}\label{app:constructionPrinciples}

In the following, we show why the ISR of the structure depicted in \cref{fig:ISRNR_Extension}(a) has the structure
  \begin{equation*}
    \mathcal{R}_S(E,H) = 
    \begin{pmatrix}
        A(E)/2 & T_{+}(E) \\
        T_{-}(E) & A(E)/2
    \end{pmatrix} \,.
    \end{equation*}
    
To show this statement, we note that there exists a block-diagonal matrix $Q$ such that  $H Q = Q H^{T}$, with
    \begin{equation*}
        Q = \begin{pmatrix}
            0 & e^{i \phi} \\ e^{i \phi}  & 0
        \end{pmatrix} \oplus I \equiv X \oplus I \,,
    \end{equation*}
    where we have enumerated the sites such that the two red sites $u,v$ are sites $1$ and $2$.
    Writing $H Q = Q H^{T}$ in block-form gives us the important relations
    \begin{align} \label{eq:symmetryRelations}
        X H_{S,S} &= \left(H_{S,S}\right)^{T} X \\
        X H_{S,\sbar} &= \left(H_{\sbar,S}\right)^{T} \label{eq:symmetryRelations2}\\
        H_{\sbar,S} &= \left( H_{S,\sbar}\right)^{T} X \label{eq:symmetryRelations3}\\
        H_{\sbar,\sbar} &= \left(H_{\sbar,\sbar} \right)^{T} \label{eq:symmetryRelations4}\, .
    \end{align}
    Equipped with these insights, we proceed by noting that the ISR over $S$ can be written as (see Lemma 4 in the Supplemental Material of Ref.~\cite{Rontgen2021PRL126180601LatentSymmetryInducedDegeneracies})
    \begin{equation}
        \mathcal{R}_S(E,H) = H_{S,S} + \sum_{k=0}^{|\sbar| - 1} a_k H_{S,\sbar} \left(H_{\sbar,\sbar} - E I \right)^{k-1} H_{\sbar,S}
    \end{equation}
    where $|\sbar|$ denotes the number of sites in $\sbar$, and with $E$-dependent coefficients $a_k$ that are obtained from the characteristic polynomial of $H_{\sbar,\sbar} - E I$.
    Now, using \cref{eq:symmetryRelations,eq:symmetryRelations2,eq:symmetryRelations3,eq:symmetryRelations4} and defining $Y \equiv H_{\sbar,\sbar} - E I$, we obtain
    \begin{align}
        &X\, \mathcal{R}_S(E,H)\, X^{-1} \\
        =& X H_{S,S} X^{-1}+ \sum_{k=0}^{|\sbar|} a_k XH_{S,\sbar} Y^{k-1} H_{\sbar,S} X^{-1} \\
        =& \left(H_{S,S}\right)^{T} + \sum_{k=0}^{|\sbar|} a_k \left(H_{\sbar,S}\right)^{T} \left(Y^{k-1} \right)^T \left( H_{S,\sbar}\right)^{T} \\
        =& \left[\mathcal{R}_S(E,H) \right]^{T} \,,
    \end{align}
    where we have used that $H_{\sbar,\sbar}$ is symmetric.
    From comparing the first and last equations, it follows that the two sites of the effective Hamiltonian must have the same energy-dependent on-site potential $A(E)/2$, while the couplings $T_{\pm}$ between them are, in general, different from each other.

\end{widetext}
\end{document}